\documentclass[%
 reprint,
 amsmath,amssymb,
 aps,
prb,
floatfix,
]{revtex4-2}

\usepackage{graphicx}
\usepackage{dcolumn}
\usepackage{bm}
\usepackage{tikz}

\usepackage{multirow}
\usepackage{array}

\usepackage[colorlinks=true,bookmarks=false,citecolor=blue,urlcolor=blue]{hyperref}

\begin{document}

\preprint{APS/123-QED}

\title{Transition between radial and toroidal orders in a trimer-based magnetic metasurface}

\author{Vladimir R. Tuz$^{1,2}$}
\author{Andrey B. Evlyukhin$^{3}$}
\author{Volodymyr I. Fesenko$^{4}$}
 \email{volodymyr.i.fesenko@gmail.com}
\affiliation{$^1$State Key Laboratory of Integrated Optoelectronics, College of Electronic Science and Engineering, International Center of Future Science, Jilin University, 2699 Qianjin St., Changchun, 130012, China}
\affiliation{$^2$School of Radiophysics, Biomedical Electronics and Computer Systems, V.~N.~Karazin Kharkiv National University, 4, Svobody Sq., Kharkiv 61022, Ukraine}
\affiliation{$^3$Institute of Quantum Optics, Leibniz Universit\"at Hannover, 30167 Hannover, Germany}
\affiliation{$^4$Institute of Radio Astronomy of National Academy of Sciences of Ukraine, 4 Mystetstv St., Kharkiv 61002, Ukraine}

\date{\today}

\date{\today}

\begin{abstract}
The change in the arrangement of magnetic dipole moments in a magnetic metasurface, due to the influence of an external static magnetic field, is discussed. Each meta-atom of the metasurface is composed of three identical disk-shaped resonators (trimer) made of magnetically saturated ferrite. To provide physical insight, full-wave numerical simulations of the near-fields and transmission characteristics of the metasurface are complemented by the theoretical description based on the multipole decomposition method. With these methods, the study of eigenmodes and scattering conditions of a single magnetic resonator, trimer, and their array forming the metasurface is performed. It is found that the magnetic dipole-based collective hybrid mode of the trimer can be gradually transformed from the radial (pseudomonopole) to azimuthal (toroidal) order and vice versa by varying the bias magnetic field strength. This is because the magnetic dipole moment of each individual disk constituting the trimer undergoes rotation as the bias magnetic field strength changes. This transition between two orders is accompanied by various patterns of localization of the electric field inside the meta-atoms. Due to the unique field configuration of these modes, the proposed metasurface can be considered for designing magnetic field sensors and nonreciprocal devices.

\end{abstract}

\pacs{78.20.Ek, 78.20.Fm, 78.67.Pt} 


\maketitle
\section{\label{intr}Introduction}

The interpretation of magnetic effects in solids has developed from two concepts \cite{Hurd_1989}. The first is the examination of \textit{microscopic} electrical currents, such as orbital currents in atoms, or magnetic dipole moments of subatomic particles related to their intrinsic angular momentum (spin). Induced magnetic moments are produced by an externally applied magnetic field, whereas spontaneous moments are present even in the absence of this field. The second concept involves the consideration of mutual interactions of the magnetic moments through ordinary dipole-dipole and quantum mechanical forces. These so-called exchange forces depend on the separation of the magnetic ions as well as their internal arrangement, leading to a variety of magnetic orders in solids. The important aspects of the study of magnetic orders regarding their nature are (i) the type of arrangement of magnetic moments and (ii) the characteristics of ordering processes associated with the corresponding phase transitions and other critical phenomena. These magnetic phase transitions can occur both from an ordered  state to a disordered state and from one magnetic order to another (e.g., the ordered magnetic moments change and become disordered at the Curie temperature; on phase transitions and critical phenomena, see Ref. \cite{stanley1971phase}). 

A typical example of the influence of order on the magnetic properties of solids is the difference between ferromagnets and antiferromagnets (these materials belong to a class of ferroics; regarding the classification of materials according to their magnetic properties, see tutorials and reference books \cite{bovca1999theoretical, coey2010magnetism, spaldin2010magnetic, cullity2011introduction, tables2003}). The order parameter of such materials is related to magnetic dipole moments and is known as magnetization. Thus, ferromagnets exhibit parallel alignment of static magnetic moments which results in the magnetization of the material even in the absence of an external bias magnetic field. In antiferromagnets, there is the presence of equal magnetic moments which are aligned in opposite directions resulting in the antiparallel moments canceling each other. This makes the total magnetic moment of antiferromagnets equal to zero. Nevertheless, the simple classification into ferromagnetic and antiferromagnetic orders cannot do justice to the rich variety of other possible ordered patterns. For instance, there are substances with molecules in which magnetization vectors have azimuthal arrangement forming closed loops \cite{Dubovik_PhysRep_1990, Spaldin_JPhys_2008, Kopaev_2009}. They are termed ferrotoroidal materials whose order parameter is called toroidization or the ferrotoroidal moment (in what follows, for brevity, the prefix `ferro' is omitted; see recent reviews on the appearance of the toroidicity in natural materials in Refs. \cite{Gnewuch_JSolidStateChem_2019, Li_DaltonTrans_2019}).

Various models are used to describe the possible orders in solids considering interactions between systems of particles and providing a framework for the study and classification of critical phenomena including phase transitions \cite{Bell_RevModPhys_2000}. In particular, the two-dimensional Ising model describes the interaction of particles that possess a magnetic dipole moment. In this model, it is supposed that the energy of a system of such particles depends on the orientation of different magnetic dipole moments relative to each other as well as to an externally applied magnetic field \cite{Newell_RevModPhys_1953}. Recently, it has been proposed to study different magnetic orders with metasurfaces \cite{Decker_OptLett_2009, Decker_PhysRevB_2009, Miroshnichenko_ACSNano_2012, Miroshnichenko_NewJPhys_2017, Lepeshov_ACSPhotonics_2018, Singh_adma_2018, yu_JApplPhys_2019, Tuz_PhysRevApplied_2020, Rahimzadegan_JOSAB_2023}. They are arrays of artificially structured subwavelength resonant particles that facilitate to reach of desirable electromagnetic functionality. Metasurfaces can maintain artificial magnetism because any desired configurations of their meta-atoms can be created \cite{Alu_2014, Kivshar_LowTempPhys_2017,evlyukhin2020bianisotropy}. In such meta-atoms, dynamic magnetic moments can arise from \textit{macroscopic} currents induced by oscillating electromagnetic fields originating from incoming radiation. 

In this regard, the use of metasurfaces for the study of toroidicity has acquired particularly significant attention. It is due to the fact that natural materials bearing the toroidal moment are scarce, and the manifestation of the toroidicity in such materials is usually weak. However, a strong toroidal response can be created in metasurfaces when their meta-atoms are properly designed. This implies the excitation of a toroidal dipole moment from a closed loop of magnetic dipoles inherent in the constituent particles forming a meta-atom. In this manner, the toroidal response is constructed in various metasurfaces whose meta-atoms are composed of either metallic \cite{Zheludev_Science_2010, Zheludev_PhysRevB_2016, Ogut_NanoLett_2012} or  dielectric \cite{Basharin_PhysRevX_2015, Tasolamprou_PhysRevB_2016, tuz_ACSPhotonics_2018} particles (for a review on the application of metasurfaces bearing toroidal moments, see Ref.~\cite{Talebi_Nanophoton_2018}).

In our previous research \cite{Xu_ADOM_2018, Dmitriev_2021, Tuz_acsanm_2020, Tuz_PhysRevB_2021, Dmitriev_OL_2021}, the general conditions for the existence of toroidal dipole moments in metasurfaces composed of nonmagnetic dielectric particles are revealed involving the multipole decomposition method and group-theoretical analysis. These conditions are checked against both numerical simulations and microwave experiments. Similar to the ferromagnetic and antiferromagnetic orders in ferroics, the toroidic and antitoroidic orders of dynamic magnetic moments are found to exist in such metasurfaces. Further development in this field may include utilizing magnetic materials in metasurfaces to implement their tunable configurations \cite{Xiao_2020, Qin_Nanophot_2022}. Specifically, these can be structures that change their dynamic magnetic order under external influence. As a significant advantage, incorporating magnetic materials into metasurface compositions can enhance magneto-optical effects and open prospects in developing advanced nonreciprocal devices \cite{Floess_2018}.

Therefore, in the present paper, our goal is  to demonstrate that in a metasurface composed of magnetic particles, the transition between different orders of magnetic dipole moments can arise under the action of an external magnetic field. These magnetic moments belong to the particles that compose the meta-atoms (here we consider meta-atoms made in the form of trimers) and their order is associated with the hybridization of the eigenmodes of the particles. These orders are the radial (pseudomonopole) and azimuthal (toroidal) arrangements of the magnetic moments (the difference between radial and azimuthal is that radial is arranged like rays that radiate from or converge to a common center while azimuthal is of or pertaining to the azimuth, i.e., in a horizontal closed loop). To describe the transition between these two orders influenced by an external static magnetic field, we utilize the multipole decomposition method developed earlier for the calculation of scattering by magnetic particles \cite{Evlyukhin_PhysRevB_2023}. The theoretical consideration is supplemented by numerical simulations performed with the commercial COMSOL MULTIPHYSICS solver for the metasurface operating in the microwave range. The simulation assumes that the particles are made from a magnetic material (ferrite), which is widely used in practice when constructing microwave devices.  

\section{Problem statement}
\label{problem}

\begin{figure}[t]
\centering
\includegraphics[width=1.0\linewidth]{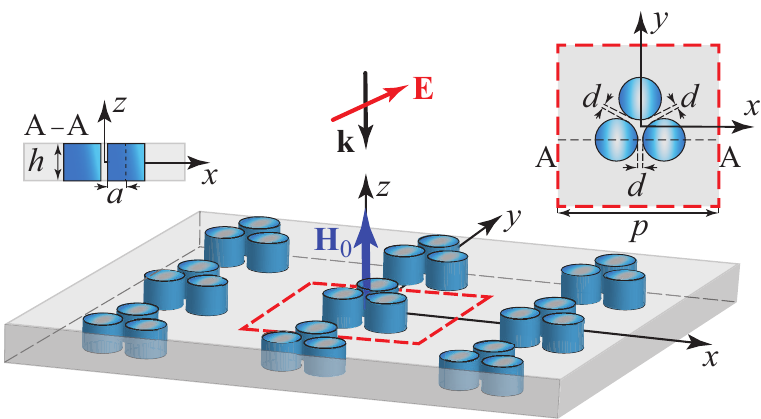}
\caption{Schematic view of a trimer-based magnetic metasurface, geometric parameters of its constituents, and the incident wave conditions. The red dashed square outlines the metasurface unit cell whose transverse and longitudinal projections are given in the insets. The blue arrow represents the direction of the externally applied static magnetic field $\textbf{H}_0$.}
\label{fig:fig1}
\end{figure} 

In what follows, we consider a metasurface composed of trimer-based meta-atoms adjusted in a two-dimensional array as shown in Fig.~\ref{fig:fig1}. The translation unit cell of this array is a square with the side size $p$. Each meta-atom consists of three identical cylindrical resonators (disks) located at the vertices of an equilateral triangle lying in the $x$-$y$ plane with their axes parallel to the $z$-axis. The distance between adjacent disks in the trimer is $d$. The radius and height of the disks are $a$ and $h$, respectively. 

We consider that the metasurface is normally irradiated by the field of a plane wave with the wave vector $\textbf{k} = \{0, 0,-k_z\}$ and wavelength $\lambda = 2\pi c/\omega$, where $\omega$ is the angular frequency and $c$ is the speed of light in vacuum. We define that the incident plane wave is either linearly or circularly polarized.

As a material for disks, we choose a widely used ferrite \cite{Pozar_book_2012}. An external static magnetic field $\textbf{H}_0=\textbf{z}_0H_0$ is applied along the $z$-axis parallel to the magnetization $\textbf{M}=\textbf{z}_0M_s$, where $\textbf{z}_0$ is the unit vector in the $z$-axis direction and $M_s$ is the saturation magnetization. Since the direction of propagation of the incident wave coincides with the bias of the external static magnetic field ($\textbf{k} \parallel \textbf{H}_0$), the problem is stated in the Faraday configuration. 

In the chosen coordinate frame and with accounting for the time factor $\exp(-i\omega t)$, ferrite is characterized by the scalar permittivity $\varepsilon$ and the second-rank permeability tensor written in the form
\begin{equation}
\hat\mu =\mu_0\left( {\begin{matrix}
   {\mu} & {i\mu_a} & {0} \cr
   {-i\mu_a} & {\mu} & {0} \cr
   {0} & {0} & {1} \cr
\end{matrix}
} \right). 
\label{eq:mu}
\end{equation}
The elements of this tensor are
\begin{equation}
\begin{split}
&\mu=1+\chi^\prime+i\chi^{\prime\prime}=\mu^\prime+i\mu^{\prime\prime},\\
&\mu_a=\chi_a^\prime+i\chi_a^{\prime\prime}=\mu^\prime_a+i\mu^{\prime\prime}_a, 
\label{eq:mu_mua}
\end{split}
\end{equation}
where $\chi^\prime=\omega_0\omega_m\left[ \omega_0^2-\omega^2(1-b^2) \right]D^{-1}$, $\omega_m=\mu_0\gamma M_s$ is the magnetization frequency, $\omega_0=\mu_0\gamma H_0$ is the Larmor frequency, $\chi^{\prime\prime} = \omega\omega_m b\left[ \omega_0^2 - \omega^2(1+b^2)\right]D^{-1}$, $\chi_a^\prime = \omega \omega_m \left[ \omega_0^2 - \omega^2(1+b^2) \right]D^{-1}$, $\chi_a^{\prime\prime} = 2b\omega^2\omega_0\omega_m D^{-1}$, $D=\left[\omega_0^2-\omega^2(1+b^2)\right]+4b^2\omega^2\omega_0^2$, $\gamma$ is the gyromagnetic ratio, $b=\mu_0\gamma\Omega/2\omega$ is a dimensionless damping constant, $\Omega$ is the linewidth of the ferromagnetic resonance. 

To be specific, commercially available $\mathsf{Al}$-doped polycrystalline yttrium iron garnet (YIG) G-400 is chosen to be a ferrite material for the disks \cite{Krupka_2021}. The properties of this material appear in the microwave range $f \in [6,12]$~GHz, $f=\omega/2\pi$, which we fix for our study. To construct the metasurface, we consider that the disks are embedded into a homogeneous medium (host) made from a material with air-like constitutive parameters. For instance, this can be a rigid foam ROHACELL-HF71 plate with $\varepsilon_h=1.09$ and $\mu_h=1$. However, other non-air-like substrates may also be used \cite{Xu_ADOM_2018}, provided that there is sufficient material contrast between the host and disks (although this paper is purely theoretical, see our experimental microwave samples of dielectric metasurfaces bearing toroidal resonances and corresponding measurement techniques in Refs. \cite{Xu_ADOM_2018, Tuz_PhysRevB_2021, Dmitriev_OL_2021}).

\section{Magnetic dipole mode of a~single ferrite resonator}
\label{single}

For the sake of completeness, we start with the description of the characteristics of a single ferrite disk-shaped resonator that is a constitutive particle of our metasurface. In particular, we are interested in the behavior of the $\textrm{HE}_{nm\ell}$ mode of this resonator, which then hybridizes into the collective mode of the trimer forming the array. In the subscripts of this mode abbreviation, the first index denotes the azimuthal variation of the fields, the second index denotes the order of variation of the field along the radial direction, and the third index denotes the order of variation of fields along the $z$-direction. Since we do not fix the exact number of the field variations along the $z$-axis, we substitute the third index with the letter $\ell$ (this notation is based on the mode nomenclature of cylindrical dielectric waveguides introduced in Ref.~\cite{Snitzer_JOSA_1961} where the subscripts $nm$ of the $\textrm{HE}_{nm}$ mode refer to the $n$th order and $m$th rank, and the rank gives the successive solutions of the boundary condition equation involving the Bessel function $J_n$).

For the HE$_{11\ell}$ mode, the fields are azimuthally dependent and the $H_z$ component is sufficiently smaller than the $E_z$ component. From the theory of dielectric resonators \cite{mongia_IntJRFMicrowave_1994}, it is known that this mode radiates like a magnetic dipole $\bf{m}$ oriented in-plane of the disk, and in terms of the Mie-theory, it is considered as a magnetic dipole (MD) mode \cite{Evlyukhin_NanoLett_2012} (hereinafter we will use this shorter notation). Within the framework of the present consideration, it is necessary to find out what changes appear in the characteristics of the MD mode when the resonator is made of ferrite influenced by an external static magnetic field. Obviously, these characteristics are determined in part by the properties of the constitutive unbounded (bulk) ferrite and the direction of application of the external static magnetic field with respect to the rotation axis of the disk-shaped resonator and the primary wave incidence.

\subsection{Magnetic subsystem conditions}
\label{sec:subsystem}

As is well known, the eigenwaves of bulk ferrite that are propagated along the magnetization direction are the left-handed (LCP) and right-handed (RCP) circularly polarized waves, which differ in the propagation constants (the Faraday effect). Near the ferromagnetic resonance, circular dichroism also takes place, which manifests itself in a significant attenuation of the LCP eigenwave as it propagates. From a phenomenological point of view, it is explained by the fact that each of those waves propagates in the medium with a different effective permeability  
\begin{equation}
\mu^{\pm}=\mu\pm\mu_a=(\mu^\pm)^\prime+i(\mu^\pm)^{\prime\prime},
\label{eq:murl}
\end{equation}
where $\mu^+$ and $\mu^-$ are related to the LCP and RCP waves, respectively. 

Moreover, in the ferrite resonator under consideration, the dominating components of the magnetic field $\textbf{H}$ of the MD mode are transverse to the bias magnetic field $\textbf{H}_0$ ($\textbf{H}\bot\textbf{H}_0$, $H_z\approx 0$). This feature and the difference in the eigenwaves propagation condition in the bulk ferrite result in the degeneracy lifting for the MD mode in the disk-shaped resonator and the corresponding eigenfrequency splitting \cite{Evlyukhin_PhysRevB_2023}. In what follows, we distinguish these split modes as $\textrm{MD}^+$ and $\textrm{MD}^-$.

\begin{figure*}[!t]
\centering
\includegraphics[width=1.0\linewidth]{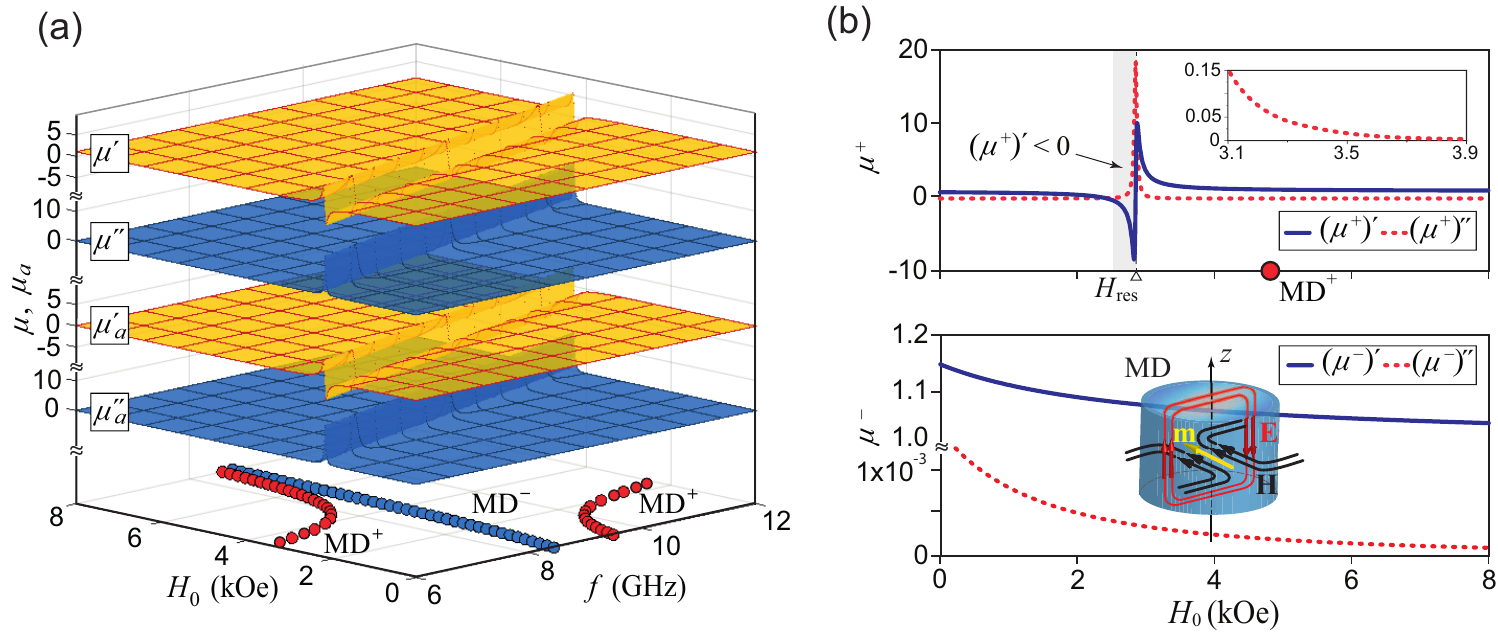}
\caption{(a) Dependencies of the real part (hot color map) and imaginary part (cold color map) of diagonal ($\mu$) and non-diagonal ($\mu_a$) components of the permeability tensor $\hat{\mu}$ of the ferrite on the frequency $f$ and bias magnetic field strength $H_0$. The colored points at the bottom plane represent the eigenfrequencies (dispersion curves) of the $\text{MD}^+$ (red) and $\text{MD}^-$ (blue) modes of the single resonator. (b)~Real and imaginary parts of the scalar effective permeabilities $\mu^+$ and $\mu^-$ as functions of $H_0$ at the fixed frequency $f=8.0$~GHz. Here, $H_{\textrm{res}} \approx\omega/\mu_0\gamma-M_s/2$ is the strength of the bias magnetic field at which the ferromagnetic resonance arises. Schematic view of the electric $\textbf{E}$ and magnetic $\textbf{H}$ fields flow of the $\text{MD}$ mode and orientation of the magnetic dipole moment $\textbf{m}$ (yellow arrow) are given in the inset. Parameters of the ferrite are $M_s=33.4$~kA/m ($4\pi M_s\sim420$~G), $\Omega\leq45$~Oe, $\tan\delta=2\times10^{-4}$, $\gamma=1.759\times10^{11}$~C/kg,  $\varepsilon=15$, whereas the radius and height of the resonator are $a=4.2$~mm and $h=6.5$~mm, respectively.}
\label{fig:fig2}
\end{figure*} 

The eigenfrequencies of the $\text{MD}^{\pm}$ modes of a ferrite disk-shaped resonator can be estimated using the following empirical relation \cite{mongia_IntJRFMicrowave_1994, Park_AP_2016}
\begin{equation}
f^\pm\approx\frac{6.324c}{2\pi a\sqrt{\varepsilon\mu^\pm+2}}\biggl[0.27+0.36\frac{a}{2h}+0.02\left(\frac{a}{2h}\right)^2\biggr].
\label{eq:freq_dl}
\end{equation}
This estimation is valid in the range $a/h \in [0.33, 5.5]$ with an accuracy of about $2\%$ for calculation of the resonant frequency for both $\text{MD}^{+}$ and $\text{MD}^{-}$ modes \cite{mongia_IntJRFMicrowave_1994}.

To show the discussed characteristics explicitly, we collect all the parameters related to the bulk ferrite and the MD mode of the ferrite resonator in Fig.~\ref{fig:fig2}. In the chosen frequency range, the ferromagnetic resonance manifests itself in the bulk ferrite when the strength of the static magnetic field $H_0$ is higher than 2~kOe. From the eigenfrequencies plotted at the bottom plane in Fig.~\ref{fig:fig2}(a), one can readily see that the $\text{MD}^{+}$ mode of the ferrite resonator is significantly perturbed by the ferromagnetic resonance and its dispersion curve has a discontinuity, whereas the $\text{MD}^{-}$ mode does not undergo any noticeable change in its dispersion. Due to circular dichroism, there is a significant attenuation of the $\text{MD}^{+}$ mode of the ferrite resonator nearby the ferromagnetic resonance frequency.

One can also notice that the  $\text{MD}^{+}$ and $\text{MD}^{-}$ modes do not degenerate in the ferrite particle even in the absence of the magnetic bias field ($H_0=0$). Previously, this peculiarity has been experimentally observed for disk-shaped resonators made of YIG as well as barium hexaferrite \cite{Popov_JApplPhys_2011, Popov_2014}. This initial mode splitting, which can reach several GHz, is not caused by residual magnetization and strongly depends on specific parameters, such as the saturation magnetization $M_s$ and the magnitude of the anisotropy field (this effect is beyond the scope of our present study; extra details on this issue can be found in Refs.~\cite{Godtmann_1967, Krupka_2021, Popov_JApplPhys_2011, Popov_2014}).

The characteristics of the effective permeabilities $\mu^\pm$ are also plotted in Fig.~\ref{fig:fig2}(b) as functions of the external magnetic field $H_0$. A particular strength of this field where the ferromagnetic resonance arises, is denoted as $H_\textrm{res}$. These curves show typical behaviors, where at the ferromagnetic resonance, the real part of $\mu^+$ undergoes changing from negative to positive values while the imaginary part reaches its maximum. The real and imaginary parts of $\mu^-$ have only a slight monotonous decrease with increasing $H_0$. 

For our further consideration, we choose the parameter space of the problem in such a way that the MD mode is in the region where the real part of $\mu^+$ is greater than zero. It is a region above the ferromagnetic resonance ($H_{\text{res}}<H_0$), where microwave devices based on low-loss ferrite materials are usually operated. A schematic representation of the flow of the electric and magnetic fields as well as the orientation of the magnetic dipole moment for the MD mode of our interest is shown in Fig.~\ref{fig:fig2}(b) just for reference.

\subsection{Resonator scattering conditions}
\label{sec:scattering}

To fully reveal the MD mode features of an individual ferrite resonator, we perform the study of scattered fields involving the multipole decomposition technique developed earlier \cite{Evlyukhin_PhysRevB_2023} to examine the scattering by magnetic particles. In general, the scattering cross-section in the multipole decomposition representation is written as \cite{Evlyukhin_PhysRevB_2023} 
\begin{eqnarray}\label{W}
    \sigma_{\rm sca}&=&\frac{k^4}{6\pi\varepsilon_0^2\varepsilon_h^2|{\bf E}_\textrm{in}|^2}|{\mathbf p}|^2+\frac{k^4}{6\pi\varepsilon_0^2\varepsilon_h^2 v^2|{\bf E}_\textrm{in}|^2}|{\mathbf m}|^2\nonumber\\
    &&+\frac{k^6}{80\pi\varepsilon_0^2\varepsilon_h^2\mu_h|{\bf E}_\textrm{in}|^2}\sum_{\beta\gamma}|Q_{\beta\gamma}|^2+\cdots
\end{eqnarray}
where $k$ is the wave number in the surrounding medium, ${\bf E}_\textrm{in}$ is the electric field amplitude of the incident plane wave. Note that in Eq. (\ref{W}) we explicitly indicate only contributions of the scatterer's electric dipole $\bf p$, magnetic dipole $\bf m$, and electric quadrupole $\hat Q$ moments. For magnetic particles, all total multipole moments are the combinations of two parts originating from the electric polarization ${\bf P}({\bf r})=\varepsilon_0(\varepsilon-\varepsilon_h){\bf E}({\bf r})$ and magnetization ${\bf M}({\bf r})=(\hat\mu-\mu_h\hat{1}){\bf H}({\bf r})$ induced by incoming radiation inside the scatterer. Here ${\bf E}({\bf r})$ and ${\bf H}({\bf r})$ are the total oscillating electric and magnetic fields at the point $\bf r$ inside the scatterer, respectively, and $\hat{1}$ is the $3\times3$ unit tensor. In general, the magnetic and electric dipole moments are given by the sums
\begin{equation}
    {\mathbf m}={\mathbf m}^\textrm{E}+{\mathbf m}^\textrm{H}, \label{ed_md}\quad {\mathbf p}={\mathbf p}^\textrm{E}+{\mathbf p}^\textrm{H}, 
\end{equation}
where the superscripts $\textrm{E}$ and $\textrm{H}$ are related to the dipole moments determined by the polarization $\mathbf P$ (electric part) and the magnetization $\mathbf M$ (magnetic part), respectively. The exact expressions for the magnetic and electric dipole moments can be found in Ref. \cite{Evlyukhin_PhysRevB_2023}. Here, for the convenience of our discussion, we present the  expressions for these dipole moments in the long wavelength approximation (LWA) which is well applicable when the size of scatterers is smaller than the incident wavelength \cite{Evlyukhin_PhysRevB_2023, Alaee2018}.
Thus, for the first-order terms of the  LWA, we have
\begin{equation}\label{eq:me}
   {\mathbf m}^\textrm{E}\approx-i\omega\frac{1}{2}\int\limits_{V_p}[{\mathbf r}\times{\mathbf P}]d{\mathbf r}, 
\end{equation}
\begin{equation}\label{eq:mh}
    {\mathbf m}^\textrm{H}\approx\frac{1}{\mu_h}\int\limits_{V_p}{\mathbf M}d{\mathbf r}    +\frac{k^2}{10\mu_h}\int\limits_{V_p}[{\mathbf r}({\mathbf r}\cdot{\mathbf M})-2r^2{\mathbf M}]d{\mathbf r},
\end{equation}
\begin{equation}\label{eq:pe}
    {\mathbf p}^\textrm{E}\approx\int\limits_{V_p}{\mathbf P}d{\mathbf r} + \frac{\mu_hk^2}{10}\int\limits_{V_p}[({\mathbf r}\cdot{\mathbf P}){\mathbf r}-2r^2{\mathbf P}]d{\mathbf r},
\end{equation}
\begin{equation}\label{eq:ph}
    {\mathbf p}^\textrm{H}\approx\frac{ik^2}{\omega\mu_h}\frac{1}{2}\int\limits_{V_p}[{\mathbf r}\times{\mathbf M}]d{\mathbf r}.
\end{equation}
Note that the second integral terms in Eqs. (\ref{eq:mh}) and (\ref{eq:pe}) are introduced as the magnetic ${\bf T}^{\rm H}$ and electric  ${\bf T}^{\rm E}$ toroidal dipole moments, respectively \cite{Gurvits2019}.

To characterize absorption in a scatterer, the absorption cross-section $\sigma_\textrm{abs}$ is used, which for magnetic particles is determined by the expression
\begin{equation}
\sigma_\textrm{abs}=\frac{k}{\varepsilon_0\varepsilon_h|{\textbf{E}_\textrm{in}|^2}}{\rm Im}\int\limits_{V_p}\{({\bf P}\cdot{\bf E}^*)+\mu_0({\bf M}\cdot {\bf H}^*)\}d{\bf r}.
\end{equation}
Note that for scatterers with real permittivity, as in our case, the contribution of $\bf P$ into the absorption cross-section is equal to zero. Therefore, the absorption is associated only with the magnetic subsystem, so that  
\begin{equation}
    \sigma_\textrm{abs}=\frac{k}{\mu_h|{\textbf{H}_\textrm{in}|^2}}{\rm Im}\int\limits_{V_p}({\bf M}\cdot {\bf H}^*)d{\bf r},
\end{equation}
where $(*)$ denotes the complex conjugation, $\textbf{H}_\textrm{in}$ is the magnetic field amplitude of the incident plane wave.

\begin{figure*}[t]
\centering
\includegraphics[width=1.0\linewidth]{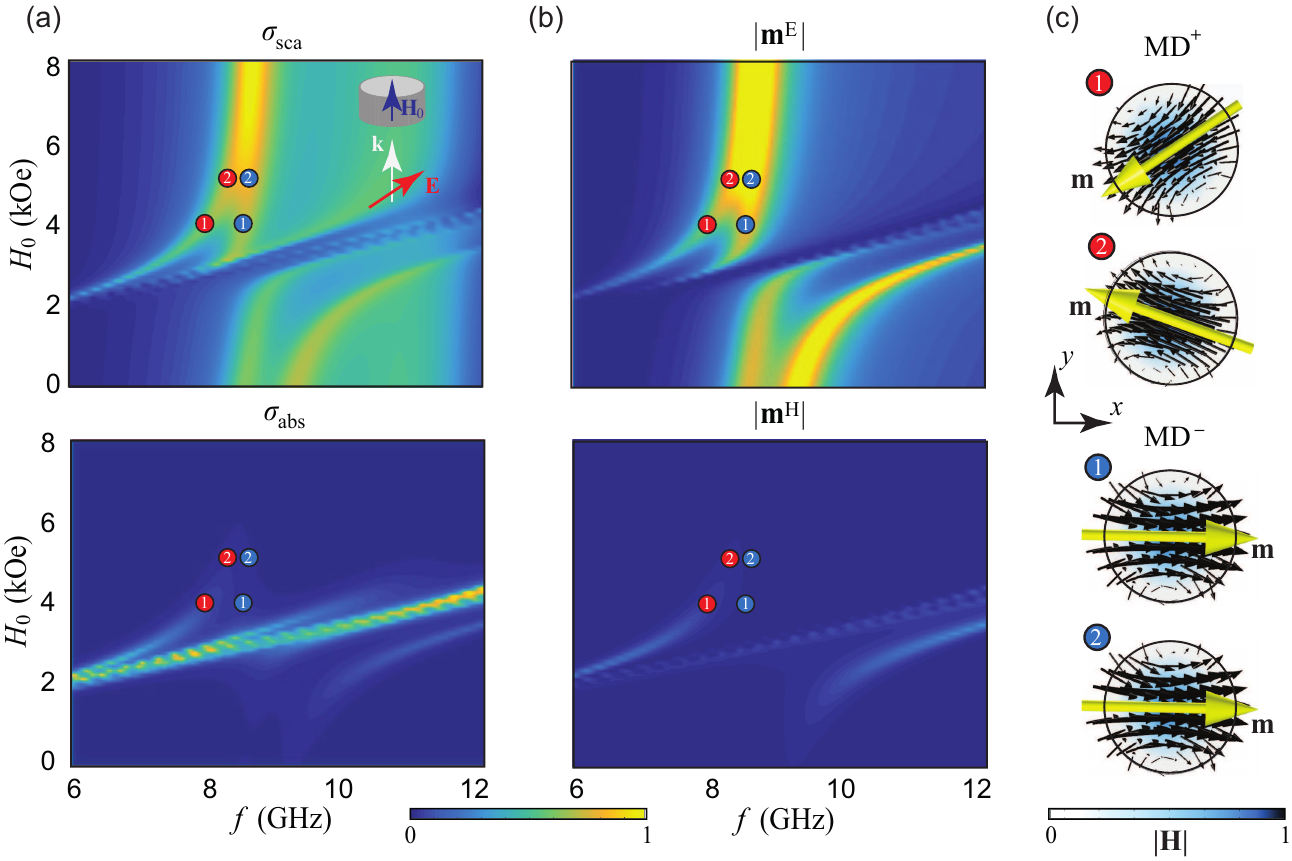}
\caption{(a) Scattering cross-section ($\sigma_\textrm{sca}$) and absorption cross-section ($\sigma_\textrm{abs}$) of a single ferrite resonator frontally irradiated by a linearly polarized plane wave ($\textbf{y}_0\parallel \textbf{E}$) as functions of the frequency $f$ and bias magnetic field strength $H_0$. All data are normalized on their corresponding maximal values. The external static magnetic field $\textbf{H}_0$ is applied in the Faraday configuration, i.e., parallel to the primary wave propagation ($\textbf{H}_0 \parallel \textbf{k}$) as shown in the inset. (b) Electric (${\bf m}^{\rm E}$) and magnetic (${\bf m}^{\rm H}$) parts of the magnetic dipole moments ($\textbf{m}={\bf m}^{\rm E}+{\bf m}^{\rm H}$). (c) Manifestation of the magnetic subsystem in the $\textrm{MD}^{\pm}$ mode of the ferrite resonator at its half-height. Here the black and yellow arrows represent the flow of the magnetic field $\textbf{H}$ and orientation of the magnetic dipole moment $\textbf{m}$, respectively. These patterns are plotted for the points \textcircled{1} and \textcircled{2} in the \{$f$-$H_0$\} parameter space depicted in panels (a) and (b). The problem parameters are the same as in Fig.~\ref{fig:fig2}.
} 
\label{fig:fig3}
\end{figure*} 

For the calculation of the scattering and absorption cross-sections as well as the  introduced dipole moments, the corresponding procedures are implemented in the RF module of the COMSOL MULTIPHYSICS solver. The realization of the computational model in this solver for both single particle and metasurface has been described earlier (see Refs. \cite{Evlyukhin_PhysRevB_2023, Kupriianov_OE_2023}), therefore here we omit all the details so as not to overburden the presentation. To reveal the scattering characteristics caused by the excitation of both $\text{MD}^{+}$ and $\text{MD}^{-}$ modes, we consider the frontal irradiation of the disk-shaped resonator by a wave with linear polarization. For definiteness, we fix that the vector $\textbf{E}$ of the incident wave is oriented along the $y$-axis. After simulation, we plot both scattering and absorption cross-sections as functions of the frequency $f$ and the external static magnetic field strength $H_0$. The implementation of expressions (\ref{eq:me})-(\ref{eq:ph}) for the dipole contributions in the solver also makes it possible to visualize the distribution of fields and the orientation of the dipole moments inside the resonator and, thus, to reveal the influence of the bias magnetic field on the MD mode behaviors. The obtained results are summarized in Fig.~\ref{fig:fig3}.

In Fig.~\ref{fig:fig3}(a), one can readily identify the manifestation of both the magnetic subsystem in its essence and the split MD resonances of the disk-shaped ferrite particle. In particular, the maximal magnitudes in the scattering and absorption cross-sections can be uniquely matched with the MD$^\pm$ dispersion curves and ferromagnetic resonance position in the \{$f$-$H_0$\} space, respectively, shown above in Fig.~\ref{fig:fig2}(a). However, since the primary wave has a linear polarization, here the areas of existence for both split MD modes experience a discontinuity in the region of ferromagnetic resonance. In fact, the MD$^\pm$ resonances disappear completely in the region where the real part of $\mu^+$ has a negative value, within which the magnetic losses are significant and the field cannot be localized inside the particle \cite{Krupka_2021}. We should note that such an effect for axially magnetized ferrite disks has been theoretically predicted \cite{Godtmann_1967} and experimentally confirmed for $\mathsf{Al}$-doped YIG disks in the microwave frequency range (see also Fig. 3 in Ref.~\cite{Krupka_2021}). This additionally verifies our results obtained.

Figure \ref{fig:fig3}(b) demonstrates the separate contributions of the electric $\textbf{m}^\textrm{E}$ and magnetic $\textbf{m}^\textrm{H}$  parts of the magnetic dipole moment $\textbf{m}$ to the scattering. One can conclude that the  $\textbf{m}^\textrm{H}$ part makes a substantial contribution near the frequency of ferromagnetic resonance, where it affects the behaviors of the MD$^+$ mode and is responsible for absorption. In the rest of the range of parameters \{$f$-$H_0$\}, the contribution of this part of the magnetic dipole moment is negligible, and the resonant scattering properties of the ferrite disk are determined mainly by the $\textbf{m}^\textrm{E}$ contribution. The apparent resonant behavior of the scattering cross-section in the range $f>10$ GHz and $H_0>3$ kOe arises due to the excitation of an electric dipole resonance, but it is not important for our present consideration.

The specific influence of the magnetic subsystem is that the overall magnetic moment $\textbf{m}$ of the MD$^+$ mode undergoes rotation as the magnetic field strength $H_0$ increases, whereas no such changes are observed for the MD$^-$ mode. The representative patterns of the magnetic field magnitude and flow as well as the orientation of the vector $\textbf{m}$ (indicated by a yellow arrow) for these split modes are shown in Fig. \ref{fig:fig3}(c). In particular, the presented pictures suggest that the rotation of the magnetic dipole moment $\textbf{m}$ for the MD$^+$ mode is clockwise, with increasing $H_0$. Further, it is of interest to find out how this rotation manifests itself in a system of several electromagnetically coupled ferrite particles, particularly, in trimers that form a metasurface.

\section{Magnetic dipole moments hybridization in a metasurface}
\label{sec:metasurface}

\begin{figure*}[t]
\centering
\includegraphics[width=1.0\linewidth]{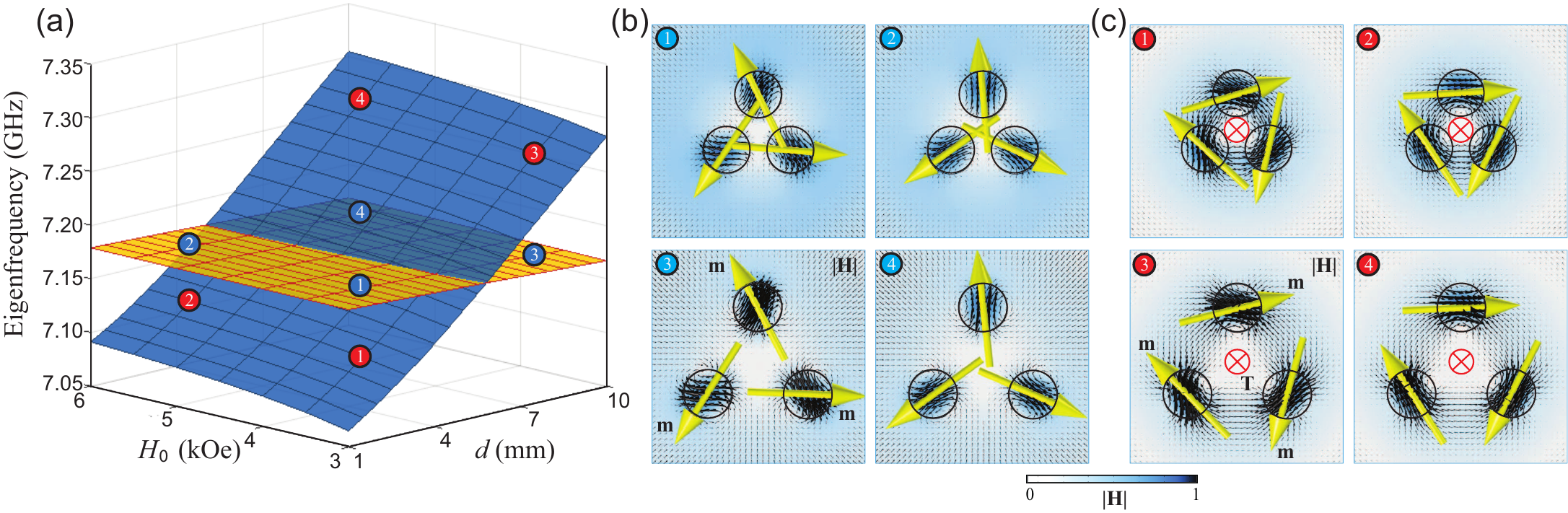}
\caption{(a) Eigenfrequencies of the radial mode (hot color map) and azimuthal mode (cold color map) of the trimer as functions of the distance $d$ and bias magnetic field strength $H_0$, and (b) and (c) schematic representations of these modes, respectively, in the trimer near-field. Here, as previously, the black and yellow arrows represent the flow of the magnetic field $\textbf{H}$ and orientation of the magnetic dipole moments $\textbf{m}$ of individual disks, respectively, whereas the red cross represents the orientation of the toroidal dipole moment $\textbf{T}$ of the trimer (it is oriented perpendicular to the $x$-$y$ plane, out from the observer). These patterns are plotted for the points \textcircled{1}--\textcircled{4} in the \{$d$-$H_0$\} parameter space depicted in panel (a). All parameters of the disks are the same as in Fig. \ref{fig:fig2}.
}
\label{fig:fig4}
\end{figure*} 

Let us now turn to the study of the spectral behaviors of our metasurface based on trimers. It is known that the electromagnetic response of the metasurface with  composite meta-atoms (clusters) containing three or more resonant particles is determined both by the properties of individual particles and, to a greater extent, by the near-field coupling between these particles within the cluster \cite{Trubin2015lattices, Maccaferri_JOSAB_2019}. This electromagnetic coupling results in existing collective hybrid modes of the cluster. Among a large variety of such collective modes, of particular interest are the so-called radial (pseudomonopole) and azimuthal (toroidal) modes \cite{Tasolamprou_PhysRevB_2016, Dmitriev_2021} that arise from the MD mode hybridization (hereinafter we denote it as the collective MD mode). These modes are not radiative, which is important from the practical viewpoint making it possible to realize strong field localization inside the system when the metasurface is operated on these modes (nonetheless, for the trimers, the pseudomonopole exhibits a nonzero magnetic octupole moment $Q^{(m)}_{yxx}$, that in some cases may unmask it \cite{Tasolamprou_PhysRevB_2016}).

We should note that the choice of a trimer-based meta-atom for the formation of our metasurface is conditioned by the fact that such a cluster provides the most straightforward rotational arrangement of particles able to support the toroidal mode \cite{Xu_ADOM_2018}. Moreover, such a cluster configuration allows one to efficiently separate the toroidal dipole from other higher-order electric and magnetic multipoles that exist in the meta-atom, compared to clusters composed of the even number of particles like a four-rod configuration \cite{Tasolamprou_PhysRevB_2016}.

\subsection{Trimer eigenmode conditions}
\label{sec:trimer}

Previously, a complete parametric study of the properties of a trimer consisting of nonmagnetic particles has been carried out to reveal the existence and excitation conditions of a collective MD mode both in a single trimer and in a metasurface based on them \cite{Tasolamprou_PhysRevB_2016,Xu_ADOM_2018, Dmitriev_2021, Tuz_acsanm_2020}. Thus, our goal here is to elucidate the effect of the external magnetic field $H_0$ on the behavior of this mode, given that this effect should be similar for its radial and azimuthal implementations. Since the existence of the collective mode of the trimer is determined by the electromagnetic coupling between its constitutive particles, which in turn depends on the distance between them, we add a variation of the parameter $d$ to this study. Here, to perform the calculation of the magnetic and toroidal dipole moments in the trimer, we utilize the secondary multipole decomposition technique \cite{Tuz_acsanm_2020}.

Figure \ref{fig:fig4} shows the change in the eigenfrequency of both the radial and azimuthal modes and the evolution of their near-field as functions of the variation of the magnetic field strength $H_0$ and the distance $d$ between the disks for a single trimer located in free space. One can see that the eigenfrequencies of both modes increase with an increase in $H_0$, whereas they have opposite trends with increasing $d$. Although both these modes of the trimer arise from the MD mode of individual disk-shaped resonators considered above, there is some difference in the arrangement of three magnetic dipole moments, each of which belongs to the corresponding particle (these magnetic moments still lie in the horizontal plane). In particular, the azimuthal mode appears from a head-to-tail arrangement of the in-plane magnetic dipole moments $\textbf{m}$ (indicated by yellow arrows located in the center of each disk) providing the toroidal dipole moment $\textbf{T}$ (indicated by a red cross located in the center of the cluster) of the trimer is oriented out-of-plane. In contrast, for the radial mode, there is the tail-to-tail (or head-to-head) arrangement of the magnetic dipole moments $\textbf{m}$ whereas the toroidal dipole moment $\textbf{T}$ is almost zero (see also Table~\ref{tab:tabA1} in Appendix~\ref{sec:appendixA}). 

\begin{figure*}[t]
\centering
\includegraphics[width=0.85\linewidth]{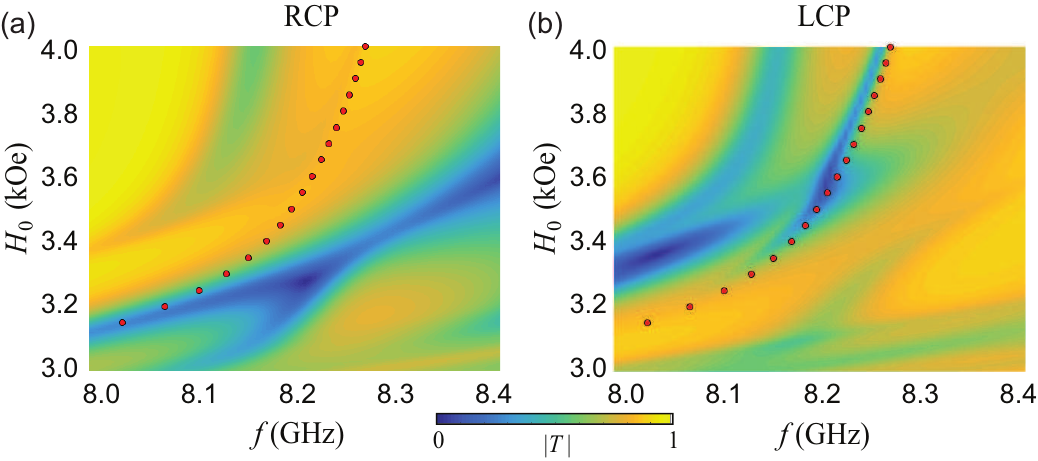}
\caption{Transmission spectra of the trimer-based magnetic metasurface illuminated by a normally incident plane wave with either (a) right-handed (RCP) or (b) left-handed (LCP) circular polarization as functions of the frequency $f$ and bias magnetic field strength $H_0$. The red points represent the eigenfrequency (dispersion curve) of the collective MD mode of the trimer. Here $p=35$~mm and $d=1$~mm, while other parameters of the problem are the same as in Fig. \ref{fig:fig2}.
}
\label{fig:fig5}
\end{figure*}

It can also be seen that as the distance $d$ between the disks increases (apparently up to some critical distance threshold when the coupling between the disks disappears), the appearance of the magnetic dipole moments $\textbf{m}$ does not change. Nevertheless, the orientation of the vectors $\textbf{m}$ changes with the variation in $H_0$. In particular, similar to that of the single particle, the rotation of the magnetic dipole moment $\textbf{m}$ in each disk of the trimer acquires a clockwise rotation, with increasing $H_0$. This change in the orientation of the moments $\textbf{m}$ can lead to the formation (breaking) of their head-to-tail arrangement for the radial (azimuthal) mode and the strengthening (weakening) of the toroidal moment $\textbf{T}$. This could potentially lead to a rearrangement of the radial mode into the azimuthal one and vice versa at a particular value of $H_0$. As we will show below, the collective effect of the metasurface makes it possible to obtain this rearrangement under relatively small variations of the external static magnetic field. 

The collective MD mode is dark for the case of frontal irradiation of the trimer ($\textbf{k} \parallel \textbf{z}_0$) since it possesses the $C_{3v}$ symmetry. Nevertheless, it can be excited in a metasurface when the trimer is deposited in a unit cell with the $C_{4v}$ symmetry. For such a symmetry reduction, it was experimentally confirmed \cite{Xu_ADOM_2018} that in a metasurface composed of trimers arranged in square unit cells, the collective MD mode can be excited by a normally incident wave with a proper polarization (see also the corresponding group-theoretical description and symmetry analysis in Ref. \cite{Dmitriev_2021}). 

\subsection{Transmission spectra peculiarities}

Before proceeding to the study of the rearrangement between the radial and azimuthal orders in the metasurface, we identify the manifestation of the collective MD mode in the transmitted spectra of the metasurface and study its polarization features. For this study, we consider that the metasurface is illuminated by a normally incident plane wave with either right-handed (RCP) or left-handed (LCP) circular polarization. Here, we have chosen circularly polarized waves to irradiate the structure, since they are eigenwaves of the longitudinally magnetized ferrite, providing independent excitation of the MD$^+$ and MD$^-$ modes for the purity of our experiment.

\begin{figure}[t]
\centering
\includegraphics[width=1.0\linewidth]{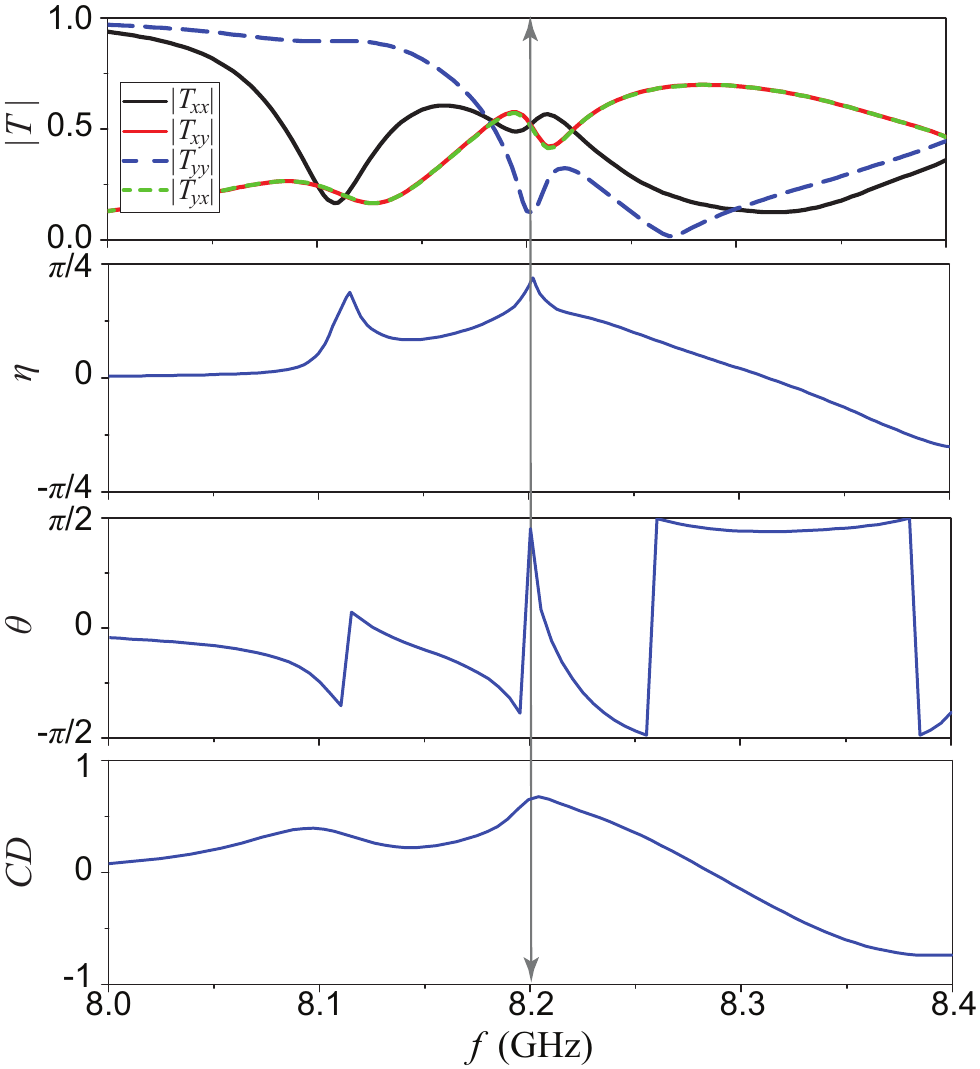}
\caption{Co-polarized ($T_{xx}$, $T_{yy}$) and cross-polarized ($T_{xy}$, $T_{yx}$, where $|T_{xy}|=|T_{yx}|$) components in terms of linear polarization, ellipticity angle ($\eta$), polarization azimuth ($\theta$), and circular dichroism ($CD$) of the transmitted waves of the trimer-based magnetic metasurface. The resonance of interest is marked by a vertical grey line ending with arrows. Here $p=35$~mm, $d=1$~mm, and $H_0=3.6$ kOe, while other parameters of the problem are the same as in Fig. \ref{fig:fig2}.
}
\label{fig:fig6}
\end{figure} 

In the solver, a single unit cell of the metasurface containing a trimer is implemented, whereas their two-dimensional arrangement is imitated by applying the periodic (Floquet) boundary conditions repeating the unit cell infinitely in both transverse directions. We dispose of the radiating and receiving ports above and below the metasurface, respectively. Just like before, after simulation, we retrieve values of the transmission coefficient $|T| = |S_{21}|$ as a function of the frequency $f$ and the strength of the external magnetic field $H_0$ for the RCP and LCP incident waves. 

The results of these calculations are summarized in Fig.~\ref{fig:fig5} where the dispersion curve of the collective MD mode is also presented. This mode of the trimer is effectively excited in the metasurface when it is irradiated with the LCP wave, while when it is irradiated with the RCP wave, only a very slight manifestation of the mode is observed in the transmission spectra. When $H_0$ increases, the corresponding resonant frequency is shifted towards higher frequencies. In contrast to the spectral feature of the LCP wave, the manifestation of the corresponding resonance in the spectra of the RCP wave is smoothed out. The overall conclusion can be made that in the region of existence of the collective MD mode, the transmission coefficient of the LCP wave approaches zero, thus the wave transmitted through the metasurface acquires RCP polarization.

These different transmissions for the RCP and LCP waves are a manifestation of the phenomenon of circular dichroism existing in the metasurface. The transmission coefficients of the RCP and LCP waves can be related directly to those of the linearly polarized waves as follows~\cite{Zhou_PhysRevB_2009}:

\begin{equation}
\begin{split}
\mathbf{T}_\textrm{circ} &= \begin{pmatrix}
T_{++} & T_{+-} \\
T_{-+} & T_{--}
\end{pmatrix} = \Lambda^{-1}\mathbf{T}_\textrm{lin}\Lambda \\
&=\frac{1}{2}
\begin{pmatrix}
T_{xx} + T_{yy} & T_{xx} - T_{yy} \\
T_{xx} - T_{yy} & T_{xx} + T_{yy}
\end{pmatrix} \\
&+\frac{i}{2}
\begin{pmatrix}
T_{xy} - T_{yx} & -T_{xy} - T_{yx} \\
T_{xy} + T_{yx} & -T_{yx} + T_{xy}
\end{pmatrix},
\end{split}  
\label{eq:rcplcp}
\end{equation}
where $\Lambda=\frac{1}{\sqrt{2}}\begin{pmatrix} 1 & 1 \\ i & -i\end{pmatrix}$ is the change-of-basis matrix. The lower signs ``$+$'' and ``$-$'' are used to describe the RCP and LCP waves, respectively, and $T_{++}$ ($T_{--}$) and $T_{+-}$ ($T_{-+}$) are their co-polarization and cross-polarization transmission coefficients. 

Using the co-polarization and cross-polarization transmission coefficients, one can determine the polarization state of the transmitted field from the following well-known formulae: \cite{Collett_1993} $\tan 2\theta = U_2/U_1$ and $\sin 2\eta = U_3/U_0$,  where $\theta$ is the polarization azimuth, $\eta$ is the ellipticity angle, and $U_j$ ($j = 0,\ldots,3$) are the Stokes parameters expressed in terms of the electric field components in the right-handed orthogonal coordinate system. According to the definition of the Stokes parameters, the ellipticity equals zero, $-\pi/4$, and $+\pi/4$ for linearly polarized, LCP, and RCP waves, respectively. In the case of $0<|\eta|<\pi/4$, the transmitted field is elliptically polarized. Then, circular dichroism can be defined as $CD=|T_{++}|^2- |T_{--}|^2$.

For the field transmitted through the metasurface, all the polarization features including circular dichroism are collected in Fig.~\ref{fig:fig6} as functions of frequency at the fixed external magnetic field strength. These figures show that the excitation of the mode of interest is related to the $E_y$-component of the incident wave ($|T_{yy}|\approx 0$), whereas, at the resonant frequency, the transmitted wave indeed becomes to be RCP polarized being influenced by significant circular dichroism. We should note that the applied external static magnetic field also breaks the time-reversal symmetry of the system leading to the property of nonreciprocity \cite{Zubritskaya_nanolett_2018, Petrucci_APL_2021}, which can be a call for future study.

\subsection{Order transition}

\begin{figure*}[!t]
\centering
\includegraphics[width=1.0\linewidth]{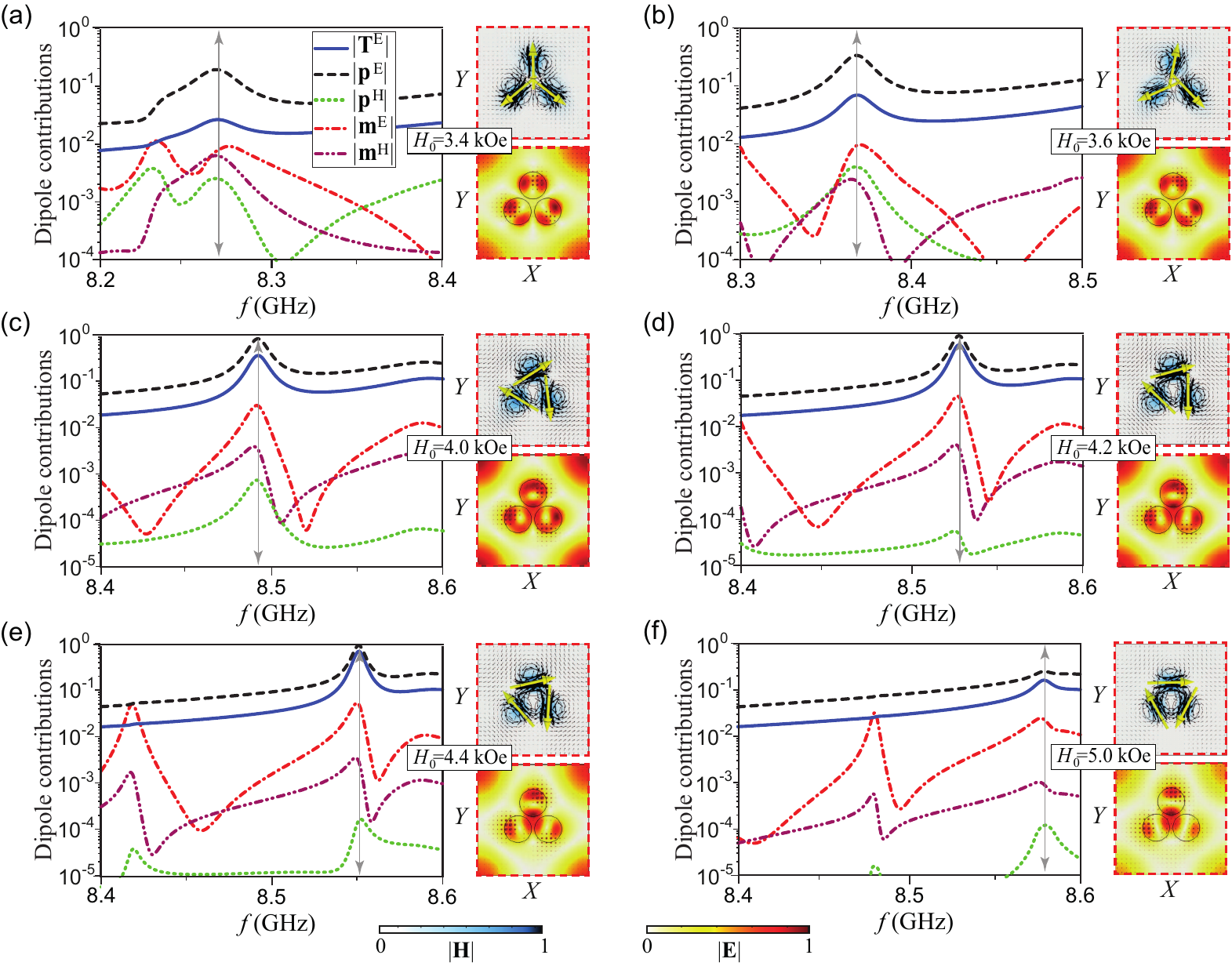}
\caption{(a-f) Contributions of the toroidal dipole, electric dipole, and magnetic dipole moments (presented on the logarithmic scale in arb. unit) to the effective scattering cross-section calculated for  the single unit cell of the trimer-based magnetic metasurface and rearrangement of the magnetic dipole moments in the trimer meta-atom with increasing the bias magnetic field strength $H_0$. The metasurface is irradiated with the LCP wave. Here, yellow, black, and red arrows represent the orientation of the individual magnetic moments of disk-shaped particles and the flow of the magnetic and electric fields within the unit cell, respectively. They are calculated at the frequencies marked by vertical grey lines ending with arrows. Here $p=35$~mm and $d=1$~mm, while other parameters of the problem are the same as in Fig. \ref{fig:fig2}.}
\label{fig:fig7}
\end{figure*}

To elucidate all conditions of the corresponding resonance, the magnitude of the toroidal dipole $\textbf{T}$, electric dipole $\textbf{p}$, and magnetic dipole $\textbf{m}$ moments of the trimer included in the metasurface unit cell are calculated for different values of $H_0$. The results are depicted in Fig.~\ref{fig:fig7}. We consider only the case when the metasurface is irradiated with the LCP wave. These are the contributions to the effective scattering cross-section, which would be in the case of scattering by the unit cell into free space (they are calculated with respect to the center of mass of the trimer using Eqs. (\ref{eq:me})-(\ref{eq:ph}), where the volume $V_s$ is replaced by the volumes of all disks forming the trimer; in our case, the centers of mass of the trimer and the unit cell coincide). The obtained spectral curves are supplemented by patterns of the magnitude distribution and flow of the magnetic and electric fields within the unit cell. These patterns are plotted in the plane fixed at the half-height of the disks (at $z=0$).

The obtained spectral curves suggest that in the selected frequency range, only the electric parts of the electric dipole ($\textbf{p}^\textrm{E}$) and toroidal dipole ($\textbf{T}^\textrm{E}$) moments make a significant contribution to the studied resonant state (we have not presented here the contribution of the magnetic part of the toroidal dipole moment ($\textbf{T}^\textrm{H}$) because of its negligible value). The small magnitude of the overall magnetic dipole moment $\textbf{m}$ is explained by the fact that, due to the specific symmetry of the trimer, the magnetic dipole moments of individual resonators are mutually compensated in the cluster. As $H_0$ increases, the resonant frequency of the collective MD mode experiences a redshift, and the magnitude of the toroidal dipole moment noticeably increases. The most remarkable feature is that with the change in $H_0$, there is also a rearrangement in the order of the magnetic dipole moments belonging to the collective MD mode.

First of all, one can see from the patterns of the electromagnetic field distributions plotted within the unit cell that the appearance of the fields in each ferrite disk closely resembles that of the MD mode of the single resonator, and for all selected values of the magnetic bias field $H_0$, there are just different manifestations of the same collective MD mode. The evolution of this collective mode consists in the fact that as $H_0$ increases, the individual magnetic moment in each disk undergoes a sequential clockwise rotation. At the extreme fixed values of $H_0=3.4$~kOe [Fig.~\ref{fig:fig7}(a)] and $H_0=5.0$~kOe [Fig.~\ref{fig:fig7}(f)], the arrangement of magnetic dipole moments in the trimer corresponds to the radial and azimuthal order, respectively, while in-between [Figs.~\ref{fig:fig7}(b-e)], this arrangement has an intermediate order.

This transformation from the radial to azimuthal order resembles a phase transition in solids and is accompanied by a change in the localization of the electric field in the central part of the cluster. Remarkably, at the stage of existence of the radial order, the electric field intensity has a local minimum in the center of the cluster and six bright hotspots on the periphery of the disks. Contrariwise, in the azimuthal order, it has a bright hotspot in the center of the cluster and three less bright hotspots on the periphery of the disks. To be specific, the amplitude of the electric field at the center of the trimer for these orders differs by a factor of ten. 

The observed feature makes it possible to realize control over the operating regimes of the trimer-based magnetic metasurface by tuning the strength of the external magnetic field. More precisely, the near-field coupling between the particles forming meta-atoms depends on the strength of the bias magnetic field, which changes the character of electromagnetic excitation of the metasurface from the radial order to the azimuthal one. Due to the distinctive field configurations of these orders, the proposed metasurface can be considered for designing magnetic field-controlled and nonreciprocal devices and magnetic field sensors (e.g., see Refs. \cite{Mund_PhysRevB_2021, Marinov_2023}).

In particular, despite the development of a large range of magnetometers, small magnetic field sensors combining high spatial and temporal resolution at room temperature remained a challenge \cite{Tumanski_book_2016}, where, among other known designs, optical sensors have the advantage of allowing non-contact measurements \cite{Lan_APLPhot_2022}. In addition to the standard parameters of the optical sensor, our metasurface has several parameters to control, namely, resonant frequency shift, change in polarization, and near-field localization order, assuming their simultaneous use can increase the sensitivity of the sensor. However, the implementation of such a sensor requires further efforts lying beyond the scope of the present consideration.

\section{Conclusions}
\label{Concl}

In conclusion, we have presented a thorough investigation of the transition of the radial (pseudomonopole) order to the azimuthal (toroidal) one in a magnetic metasurface conditioned by the influence of an external static magnetic field. The metasurface with trimer-based clusters arranged in a two-dimensional array with the translation unit cell based on identical disk-shaped ferrite resonators (made of $\mathsf{Al}$-doped YIG material) has been considered. To provide physical insight, the numerical simulations have been supplemented by an analytical description of the eigenfrequencies and scattering characteristics of the individual ferrite resonator and trimer based on them. The evolution of the collective mode ordering in the trimer with changing the bias magnetic field has been considered. It was found that the collective magnetic dipole mode of the trimer can be gradually transformed from the radial to toroidal order by increasing the bias magnetic field strength. 

We believe that the proposed design of the metasurface can be helpful in practical applications due to the possibility of easily switching the nature of the excitation between different magnetic orders and conditions of the electric field localization inside the system. Since electromagnetic effects obey the principle of duality, the presented phenomena should manifest themselves in metasurfaces containing particles made of optically active materials with tensor-valued permittivity. In this case, one can expect the rearrangement of the orders generated by a collective electric dipole mode (in particular, this can originate from the EH$_{11\ell}$ mode of a disk-shaped resonator). This may be important in terms of scaling such metasurfaces to a higher frequency range.

\section*{Acknowledgments}
V.R.T. is grateful for the hospitality and support from Jilin University, China. A.B.E. thanks funding support from the Deutsche Forschungsgemeinschaft (DFG, German Research Foundation) under Germany’s Excellence Strategy within the Cluster of Excellence PhoenixD (EXC 2122, Project ID 390833453).

\bigskip

\appendix
\section{Symmetry-adapted linear combination (SALC) approach}
\label{sec:appendixA}
\setcounter{table}{0}
\renewcommand{\thetable}{A}

\begin{figure}[t]
\centering
\includegraphics[width=0.5\linewidth]{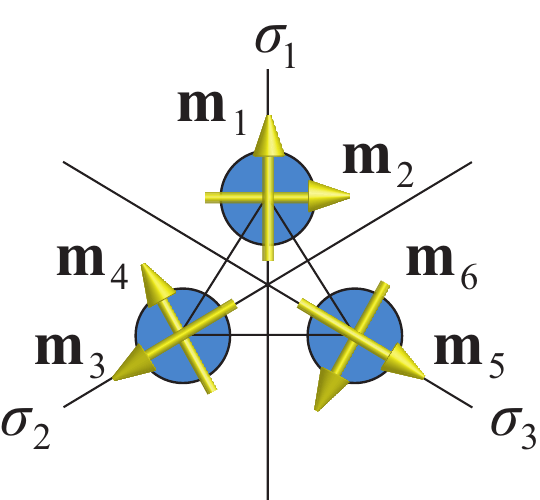}
\caption{Symmetry adapted basis for the trimer with the $C_{3v}$ symmetry.}
\label{fig:figB1}
\end{figure}

The chosen basis for dipole magnetic moments is shown in Fig. \ref{fig:figB1}. The unit vectors ${\bf m}_i$ with odd and even index $i$ are oriented parallel and perpendicular to the planes of symmetry $\sigma$, respectively. One can calculate SALCs using the projection operators \cite{Bradley_book_2009}:
\begin{equation}
{\bf m}_{kj}^{\Gamma} \!=\! \dfrac{1}{g} \sum_{R}\chi_{kj}^{\Gamma}(R)R{\bf m}_i,
\label{eq:22}
\end{equation}
where the summation is performed with respect to the elements $R$ of the $C_{3v}$ group. These elements are $e$, $C_3$, $C_3^{-1}$, $\sigma_{1}$, $\sigma_{2}$, and $\sigma_{3}$ (see Ref. \cite{Dmitriev_2021}). $\Gamma$ is a representation $A$ or $B$ of the $C_{3v}$ group, $\chi_{kj}^{\Gamma}(R)$ is the $kj$-th component of the irreducible representation (IRREP) $\Gamma$ of the element $R$, $R{\bf m}_i$ is a basis element ${\bf m}_i$ transformed by the operator $R$, and $g$ is the order of the group (in our case, $g=6$).

The normalized eigenvectors of the trimer are presented in the third column of Table~\ref{tab:tabA1}. Corresponding eigenmodes are also identified performing the full-wave simulation in Ref. \cite{Tasolamprou_PhysRevB_2016}.

%

%
\begin{table}[!h]
\label{tab:tabA1}
\begin{center}
\caption{Azimuthal (toroidal) and radial (pseudomonopole) eigenmodes of a trimer with the $C_{3v}$ symmetry in terms of vectors ${\bf m}_i$}
{
\begin{tabular}{c@{\hspace{4mm}}c@{\hspace{4mm}}c@{\hspace{4mm}}
} \hline
\\
IRREP & Eigenmode image & Description \\  
\\  
\hline
\hline
\raisebox{7ex}{$A$} 
& \includegraphics[height=2.5cm]{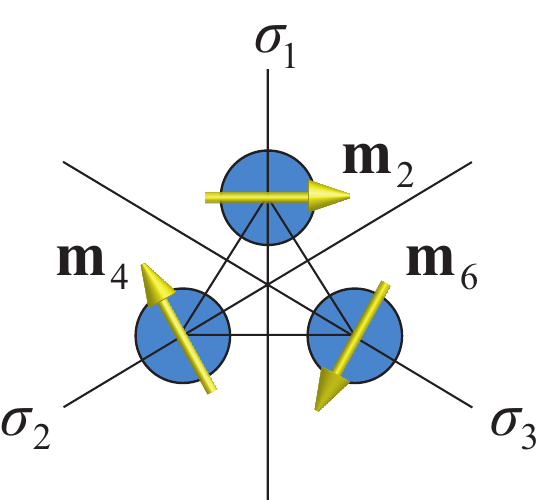}
& \raisebox{7ex}{
    $\begin{array}{c}
    \textrm{Azimuthal mode} \\
    \dfrac{{\bf m}_2+{\bf m}_4+{\bf m}_6}{\sqrt{3}}
    \end{array}$
}
\\
\hline
\\
\raisebox{7ex}{$B$}
& \includegraphics[height=2.5cm]{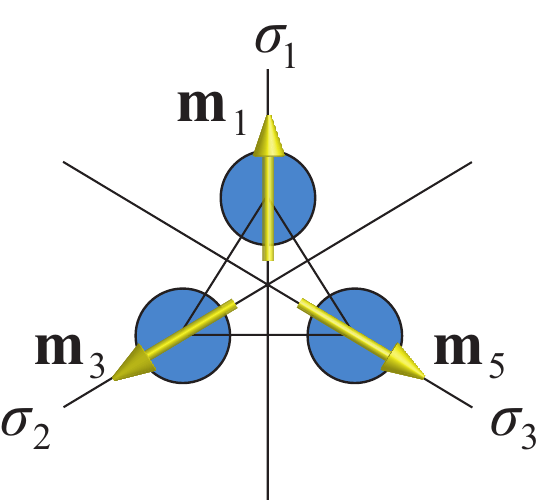}
& \raisebox{7ex}{
    $\begin{array}{c}
    \textrm{Radial mode} \\
    \dfrac{{\bf m}_1+{\bf m}_3+{\bf m}_5}{\sqrt{3}}
    \end{array}$
}
\\
\hline	
\hline
\end{tabular}
}
\end{center}
\end{table}

\bigskip

\bibliography{toroid}

\end{document}